\documentclass{emulateapj}

\usepackage{subfloat}

\newcommand{\HI}{\hbox{{\rm H}\kern 0.2em{\sc i}}}

\shorttitle{A Phase-space View of VIVA Galaxies}
\shortauthors{Yoon et al.}

\begin{document}

\title{A history of {\HI} stripping in Virgo: a phase-space view of VIVA galaxies}


\author{Hyein Yoon\altaffilmark{1}}
\author{Aeree Chung\altaffilmark{1,2,3}}
\author{Rory Smith\altaffilmark{1}}
\author{Yara L. Jaff\'{e}\altaffilmark{4}}

\affiliation{\altaffilmark{1}Department of Astronomy, Yonsei University, 50 Yonsei-ro, Seodaemun-gu, Seoul 03722, Korea; \\ hiyoon@galaxy.yonsei.ac.kr; achung@yonsei.ac.kr\\
\altaffilmark{2}Yonsei University Observatory, Yonsei University, Seoul 03722, Korea\\
\altaffilmark{3}Joint ALMA Observatory, Alonso de Cord\'{o}va 3107, Vitacura, Santiago, Chile\\ \altaffilmark{4}European Southern Observatory, Alonso de C\'{o}rdova 3107, Vitacura, Casilla 19001, Santiago, Chile}

\begin{abstract} 

We investigate the orbital histories of Virgo galaxies at various stages of {\HI} gas stripping. In particular, we compare the location of galaxies with different {\HI} morphology in phase space. This method is a great tool for tracing the gas stripping histories of galaxies as they fall into the cluster. Most galaxies at the early stage of {\HI} stripping are found in the first infall region of Virgo, while galaxies undergoing active {\HI} stripping mostly appear to be falling in or moving out near the cluster core for the first time. Galaxies with a severely stripped, yet symmetric, {\HI} disks are found in one of two locations. Some are deep inside the cluster, but others are found in the cluster outskirts with low orbital velocities. We suggest that the latter group of galaxies belong to a ``backsplash" population. These present the clearest candidates for backsplashed galaxies observationally identified to date. We further investigate the distribution of a large sample of {\HI}-detected galaxies toward Virgo in phase space, confirming that most galaxies are stripped of their gas as they settle into the gravitational potential of the cluster. In addition, we discuss the impact of tidal interactions between galaxies and group preprocessing on the {\HI} properties of the cluster galaxies, and link the associated star formation evolution to the stripping sequence of cluster galaxies.

\end{abstract}


\section{Introduction} \label{sec:intro}

Gas stripping from galaxies is one of the main drivers of galaxy evolution \citep[e.g.][]{abadi99,vollmer01,roediger05,boselli06}. Among a number of externally driven gas-stripping mechanisms, ram pressure due to the surrounding medium has been suggested as an effective way to remove interstellar gas from a galaxy in a relatively short time \citep{gunn72, abadi99}. This is particularly true in the cluster environment. Around clusters with a mass of a few 10$^{14}$~$M_{\odot}$ or higher, galaxies tend to radially fall into the deep potential well, experiencing dynamical as well as thermal pressure due to the dense intracluster medium (ICM). Indeed, in nearby clusters, many clear cases of ram-pressure stripping have been caught in action by high-resolution {\HI} imaging studies \citep{chung07,chung09}.

Ram pressure is expected to be strongest near the pericenter of a galaxy's orbit, as both the ICM density and the orbital velocity are highest at this location \citep{gunn72}. This prediction is supported by the general trend that the fraction of {\HI}-stripped galaxies increases with decreasing projected distance from the cluster core \citep{solanes01, boselli06, jaffe15}. However, there is a large scatter about this trend, and severely gas-stripped galaxies are found at a range of environmental densities, including cluster outskirts, in real and simulated galaxy clusters alike \citep{solanes01,tbvg07}. 

There are several possible explanations, including (1) merging of subgroups into the main cluster, which may generate non-static ICM \citep{shibata01}, (2) preprocessing within infalling subgroups and/or filamentary structures \citep{fujita04,jaffe16}, or (3) backsplashing of galaxies \citep{gill05}. Hence, in order to study the detailed histories of gas stripping and star formation of galaxies in the cluster environment, it is important to probe not only the current surroundings of galaxies but also which environments galaxies have gone through in the past, i.e. their possible trajectories before arriving where they are now. 

Therefore, it is desirable to characterize in the 3D distribution of galaxies in galaxy clusters. However, this is not trivial, due to the proper motions of cluster members. For late-type galaxies, which are critical for studying the role of gas stripping in galaxy evolution, distance measurements can be attempted using Hubble constant-independent distance indicators like the Tully$-$Fisher relation \citep{tully77}. But in practice, the application of the Tully$-$Fisher can be limited since both neutral and ionized gas tracers such as {\HI} and H$\alpha$ are often severely truncated in galaxies in high-density regions \citep{koopmann04b,chung09,taylor13}. Moreover, tracing a galaxy's trajectory from a single snapshot is even more difficult, if not impossible. 

Instead, it has been shown that the position-velocity space (a.k.a. phase space) can provide a useful tool to probe the orbital histories of galaxies \citep{oman13, hernandez14, haines15, jaffe15}. For example, using two observables, the projected clustocentric distance and the line-of-sight velocity with respect to the cluster mean, \cite{oman13} show that it is feasible to identify regions where galaxies are at different stages of their orbit within the cluster, such as ``infalling," ``backsplashing," or within a ``virialized" region \citep[see also][]{mahajan11}, based on their location in projected phase space. \cite{jaffe15} showed that, as a result of these different orbital stages, a clear signature of ram pressure could be detected in the phase-space distribution of {\HI}-detected cluster galaxies. Furthermore, \cite{oman16} present the timescales of quenching in star formation in cluster orbits, by using probability distributions between infall time and each position of phase space.

As one of the most diffuse and extended forms of interstellar gas, {\HI} often provides a great tool to diagnose the stripping status of a galaxy \citep{chung07,chung09}. Hence, phase-space diagrams of cluster galaxies, identified at various stages of {\HI} stripping, can be a good way to trace back the trajectories that galaxies followed while undergoing ram-pressure stripping. This is especially the case in nearby clusters, where the time since gas stripping started taking place actively can be inferred by {\HI} peculiarities, like one-sided long tails and extraplanar gas \citep{chung07}. Therefore, the combined information from high-resolution {\HI} data and a phase-space diagram can be an ideal tool for understanding when and where galaxies lose their gas.

In this work, we thus attempt to probe the detailed orbital histories of gas-stripped galaxies, using a phase-space analysis. We specifically focus on Virgo cluster galaxies with high-resolution {\HI} imaging data, which is an excellent indicator of the stripping stage.

This paper is organized as follows. In Section~\ref{sec:sample}, we introduce the sample. In Section~\ref{sec:phase_gen}, we describe the detailed prescriptions of how a phase-space diagram for our sample is constructed. In Section~\ref{sec:phase_viva}, we probe the locations of galaxies at various stages of {\HI} stripping. In Section~\ref{sec:discuss}, we discuss the orbital histories of {\HI}-stripped galaxies, and the impact of main/subcluster structures on gas stripping. In Section~\ref{sec:sum}, we summarize the results. Throughout this work, we adopt a distance of 16.5~Mpc (or $m-M=31.1$) to the Virgo cluster \citep{mei07}.

\begin{deluxetable*}{cccrcl}[h]
\tablecaption{The VIVA sample separated into various stages of gas stripping \label{tab:table1}}
\tablewidth{0pt}
\tablehead{
\colhead{Class} & \colhead{$D_{\rm\tiny {\HI}}/D_{\rm opt}$} & \colhead{$D_{\rm\tiny {\HI}}/D_{\rm opt}$} & \colhead{$def_{\rm\tiny {\HI}}$} & \colhead{$def_{\rm\tiny {\HI}}$} & \colhead{Galaxy} \\
\colhead{} & \colhead{(Range)} & \colhead{(Median)} & \colhead{(Range)} & \colhead{(Median)}  &\colhead{}\\
\colhead{(1)}&\colhead{(2)}&\colhead{(3)}&\colhead{(4)}&\colhead{(5)}&\colhead{(6)}
}
\startdata
Class~0 & 0.83 $\sim$ 4.19 & 1.90 & $-$0.81 $\sim$ 0.38 & 0.13 & \object{NGC 4189}, \object{NGC 4222}, \object{NGC 4321}, \object{NGC 4383},\\
(13~galaxies) &&&&& \object{NGC 4532}, \object{NGC 4536}, \object{NGC 4561}, \object{NGC 4567},\\
&&&&& \object{NGC 4568}, \object{NGC 4651}, \object{NGC 4713}, \object{NGC 4772},\\
&&&&& \object{NGC 4808}\\
Class~I & 1.20 $\sim$ 2.21 & 1.37 & $-$0.43 $\sim$ 0.41 & 0.02 & \object{NGC 4254}, \object{NGC 4294}, \object{NGC 4299}, \object{NGC 4351},\\
(7~galaxies) &&&&& \object{NGC 4396}, \object{NGC 4535}, \object{NGC 4698}\\
Class~II & 0.39 $\sim$ 1.53 & 0.76 & 0.12 $\sim$ 1.16 & 0.76 & \object{NGC 4298}, \object{NGC 4302}, \object{NGC 4330}, \object{NGC 4388},\\
(10~galaxies) &&&&& \object{NGC 4402}, \object{NGC 4424}, \object{NGC 4501}, \object{NGC 4522}, \\
&&&&& \object{NGC 4654}, \object{NGC 4694}\\
Class~III & 0.20 $\sim$ 0.70 & 0.43 & 0.82 $\sim$ 2.25 & 1.50 & \object{NGC 4064}, \object{NGC 4293}, \object{NGC 4405}, \object{NGC 4419},\\
(10~galaxies) &&&&& \object{NGC 4457}, \object{NGC 4569}, \object{NGC 4580}, \object{NGC 4606},\\
&&&&& \object{NGC 4607}, \object{IC 3392}\\
Class~IV & 0.57 $\sim$ 1.04 & 0.73 & 0.51 $\sim$ 1.17 & 0.79 & \object{NGC 4192}, \object{NGC 4216}, \object{NGC 4380}, \object{NGC 4394},\\
(8~galaxies) &&&&& \object{NGC 4450}, \object{NGC 4548}, \object{NGC 4579}, \object{NGC 4689}\\
\enddata
\tablecomments{(1) Class, (2) the range of {\HI} isophotal diameter-to-optical $B$-band diameter (more detailed definitions can be found in the text), (3) the median of column (2), (4) the range of {\HI} deficiency, (5) the median of column (4), and (6) galaxies in each subclass. Columns (2) and (4) are taken from \cite{chung09}.}
\end{deluxetable*}

\section{Sample selection} \label{sec:sample}

\subsection{Classification of gas-stripped VIVA galaxies}

The VLA Imaging survey of Virgo galaxies in Atomic gas \citep[VIVA I;][]{chung09} provides an ideal sample for this study. The VIVA sample is spread throughout the cluster, out to a clustocentric distance of $\sim$3~Mpc \citep[2 $\times$ $R_{200}$;][]{mclaughlin99,ferrarese12}. Hence, it encompasses a range of environmental densities and contains galaxies caught in a variety of stages of {\HI} stripping. For this sample, combining location in phase space, with {\HI} morphology can provide a wealth of information on a galaxy's evolutionary history.

High-resolution {\HI} imaging is a powerful tool for probing environmental processes. By studying peculiarities in {\HI} morphology in combination with multiwavelength data, the key mechanism impacting individual galaxies can be identified \citep[e.g.][]{vollmer09, abramson11}.  In this work, we are particularly interested in the cases that are undergoing, or have gone through, {\HI} stripping due to the surrounding ICM. For this, we utilized {\HI} deficiency, relative extent to the stellar disk, and morphology, which are the key parameters in diagnosing various stages of gas stripping. We left out dwarfs for which {\HI} properties vary quite widely among the field population, and only considered 48~disky galaxies out of 53~VIVA galaxies. Many display peculiarities in their {\HI} morphology, which are commonly associated with {\HI} stripping, such as asymmetry or truncation of their gas disk within the radius of the stellar disk.

In total, 35~galaxies have been selected to be undergoing gas stripping. These can be categorized roughly into four groups, as listed below:

\begin{enumerate}
\item[(i)] Class~I: One-sided {\HI} feature, such as a tail, and no truncation of the {\HI} disk within the relatively symmetric stellar disk; range of {\HI} deficiencies shown, but overall comparable to those of field galaxies.

\item[(ii)] Class~II: A highly asymmetric {\HI} disk, with one-sided gas tails, extraplanar gas, and/or {\HI} disk truncation on at least one side of the stellar disk, as if the {\HI} disk has been compressed; quite deficient in {\HI}, with an average of only $\sim$17\% of the typical {\HI} mass of a field counterpart.

\item[(iii)] Class~III: A symmetric, but severely truncated {\HI} disk; extremely deficient in {\HI} with an average of $<$ 4\% of the {\HI} mass of a field galaxy counterpart.

\item[(iv)] Class~IV: A symmetric {\HI} disk with marginal truncation within the radius of the stellar disk; lower {\HI} surface density than the other subclasses; quite deficient in {\HI}, with on average $\sim$15\% of the {\HI} mass of a field galaxy counterpart. 
\end{enumerate}

The remaining~13 galaxies are one of the following cases: (1) quite symmetric, not truncated, not highly deficient in {\HI} like normal field spirals, or (2) extremely rich and extended in {\HI} with no one-sided feature, or (3) asymmetric but not truncated in {\HI} with clear signs of tidal interaction. These are therefore the cases showing no definite signs of gas stripping due to the ICM, and we have grouped them as Class~0.

\begin{figure*}
\epsscale{1.2}
\plotone{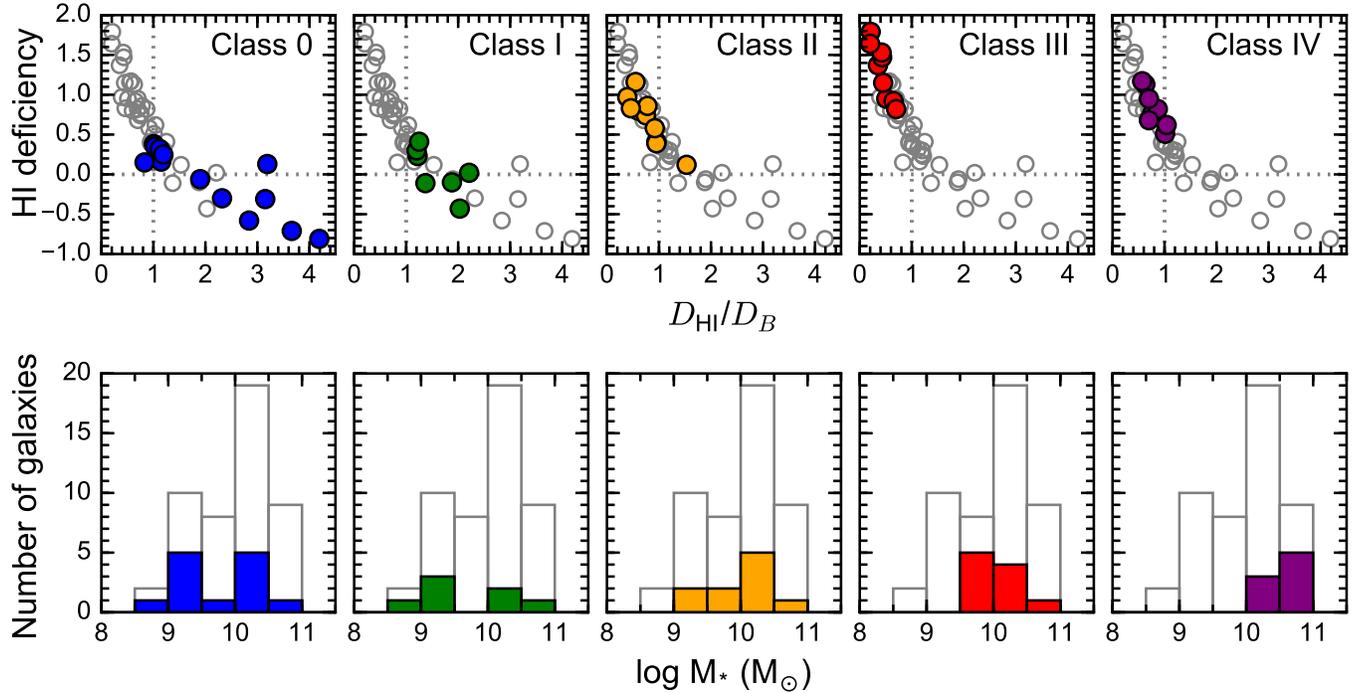}
\caption{{\HI} deficiency vs. {\HI}-to-optical $B$-band size of VIVA galaxies adopted from \cite{chung09}. In each panel, galaxies of each subclass are indicated by a particular color. Blue is Class~0, green is Class~I, yellow is Class~II, red is Class~III, and purple is Class~IV. The dotted horizontal line represents $def_{\rm\tiny {\HI}}=0$, i.e. the same {\HI} mass as a field counterpart with the same size. The vertical dotted line indicates where the {\HI} and stars have equal diameters. The definitions of {\HI} and optical $B$-band diameter can be found in Section~\ref{sec:sample}. In the lower row, the stellar mass distribution of each class is also presented, where galaxies are found to be randomly distributed among each class. \label{fig:fig1}}
\end{figure*}

The characteristics of the five classes, and the differences among them, are quantitatively presented in Table~\ref{tab:table1} and Figure~\ref{fig:fig1}. $D_{\rm\tiny {\HI}}/D_{\rm opt}$ is the relative {\HI} extent. $D_{\rm\tiny {\HI}}$ is the {\HI} disk size measured at the radius where the azimuthally averaged {\HI} surface density drops to 1~$M_{\odot}$/pc$^{2}$ \citep{chung09}. $D_{\rm opt}$ is the isophotal diameter measured at 25th mag~arcsec$^{-2}$ in $B$ band \citep[RC3;][]{devaucouleurs91}. {\HI} deficiency, $def_{\rm\tiny {\HI}}$, is the measure of how deficient one galaxy is in {\HI} compared to its field counterpart \citep[$def_{\rm\tiny {\HI}}=<$~log~M$_{\rm\tiny {\HI}}>-$~log M$_{\rm\tiny {\HI}}$;][]{haynes84}. In this work, we adopt the values from \cite{chung09}, which takes account of the galaxy size, but not the morphology. 

The stellar mass distribution of the sample is also shown in the lower row of Figure~\ref{fig:fig1}. Overall, each class seems to be rather randomly (but not uniformly) distributed, while Classes~III and IV show a higher fraction of massive galaxies. Considering the sensitivity of the VIVA study, this trend does not necessarily reflect stronger stripping of lower mass galaxies. Rather, this seems to originate from the sample selection strategy of the VIVA study. Compared to the previous VLA {\HI} imaging study by \cite{cayatte90}, the VIVA survey included more low-mass galaxies at a wider range of clustocentric distances. This naturally led the sample to contain quite a high fraction of less massive galaxies that are not extremely {\HI} deficient. However, since the stellar mass distribution of the most gas-poor group (Class~III) is not significantly different from that of Classes~0 to $\sim$III, the conclusions of our study are not expected to be significantly affected by stellar mass.

The {\HI} properties of each class likely represent distinct types of ICM$-$ISM interactions or various stages of ram-pressure stripping. {\HI} gas is usually extended well beyond the stellar disk in the low-density environment \citep{warmels88,broeils94}, and hence it is the first component to get affected by external mechanisms. While tidal interactions are expected to strip material from a galaxy in a more symmetric manner, ICM winds tend to push the interstellar medium to one side during the first infall of the galaxy. As the galaxy continues to fall into the cluster, the strength of the ram pressure increases. Some, or most, of the gas is stripped from the galaxy, leaving a tail or extraplanar gas in some cases. As the gas gets stripped, the gas disk gets increasingly truncated within the radius of the stellar disk, starting from one side of the gas disk. 

Based on the predictions above, Class~0 galaxies are likely to represent prestripping cases, while Classes~I and II might be good examples of galaxies undergoing early and active ram-pressure stripping, some of which might potentially be experiencing tidal interactions at the same time. If a galaxy plunges quite deeply inside of the cluster during the first infall, the extraplanar features may disappear after the first core crossing, once the stripped gas has left the vicinity of the galaxy. After some time, the remaining gas may resettle into the disk \citep{vollmer04}, resulting in a severely truncated and somewhat symmetric gas disk, like the Class~III galaxies. 

Considering the limited bandwidth and the sensitivity of VIVA, we cannot completely rule out the possibility of intracluster {\HI} streams near Class~III galaxies (e.g. the {\HI} plume found near \object{NGC 4388} by \cite{oosterloo05}, which was missed in VIVA). However, if Class~III galaxies are the examples of past stripping as we propose, the presence of {\HI} streams next to these galaxies is unlikely, since the gas lost by galaxies will be dispersed/evaporated with time. Especially after a galaxy's core crossing, such gas streams are likely to lose the association with galaxies even if they survive for a while. Indeed, the intracluster atomic hydrogen gas found to date within a $\sim$1~Mpc radius around \object{M87} is always associated with galaxies at early to active ram-pressure stripping stages (e.g. \object{NGC 4254} by \cite{haynes07}; \object{NGC 4388} by \cite{oosterloo05}) in contrast the galaxies that have lost the gas quite a while ago. Although all of these examples are based on limited observations, and even if finding {\HI} streams next to Class~III galaxies is still possible, their {\HI} morphology within the stellar disk, which is highly truncated and symmetric, clearly implies that they are at later stripping stages than Class~II galaxies.

Meanwhile, Class~IV galaxies might be the result of mechanisms that can affect the entire disk on longer timescales compared to ram-pressure stripping. This class is comparable to Class~II galaxies in both the relative {\HI} extent and deficiency, but the {\HI} gas morphology is very distinct. Unlike Class~II galaxies, Class~IV galaxies have a quite symmetric {\HI} disk that is marginally truncated within the stellar disk. Hence, processes that take place throughout the entire disk simultaneously on long timescales (unlike in Class~III objects), such as thermal evaporation or starvation, might be responsible for objects in this class.

Among the total VIVA sample of $\sim$50~galaxies, this subsample of 35~galaxies shows the clearest signs of an ICM$-$ISM interaction, yet we do not rule out the possibility that these galaxies might be simultaneously gravitationally perturbed.

\subsection{{\rm H}\kern 0.2em{\sc i} control sample in and around Virgo}

\begin{figure*}
\epsscale{0.8}
\plotone{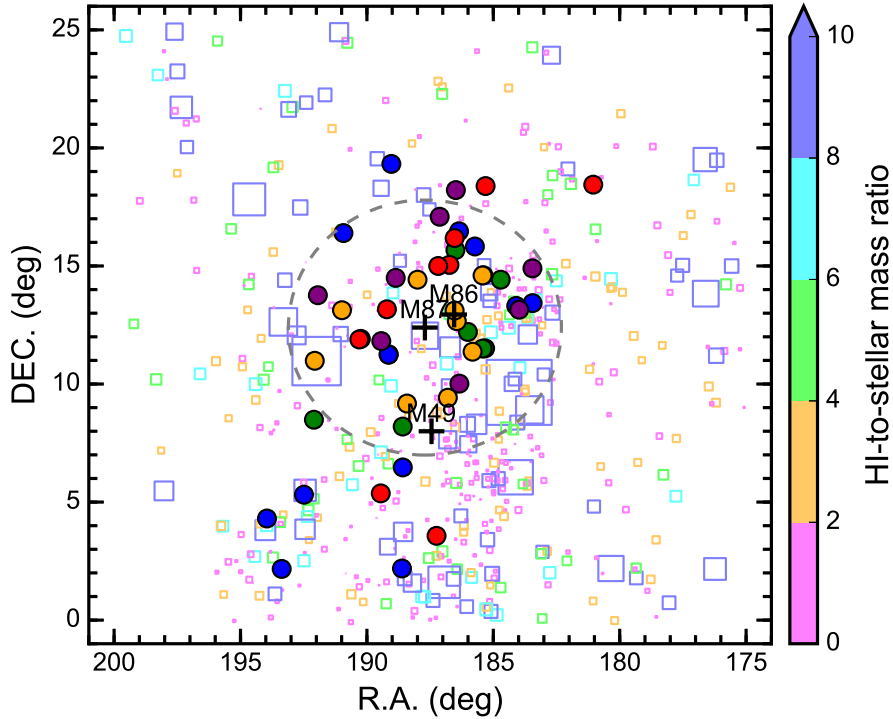}
\caption{Spatial distribution of {\HI}-detected galaxies in and around the Virgo cluster. Filled circles are the sample from this study, with the same color scheme as in Figure~\ref{fig:fig1}. Open squares are the reference sample, collected from the ALFALFA $\alpha$.100 ``A-grids" catalog. The symbol color (see color bar) and size indicate the {\HI}-to-stellar mass ratio. The dashed circle indicates the $R_{200}$~($\sim$5.$^{\circ}$4) of Virgo.
 \label{fig:fig2}}
\end{figure*}

In addition to the various subclasses of the VIVA sample, the distribution of {\HI}-detected galaxies in the Virgo cluster and the surrounding environments in phase space is used as a reference, as shown in Figure~\ref{fig:fig2}. Thanks to the recent release of the ALFALFA ``A-grids" catalog ($\alpha$.100 ``A-grids")\footnote{The 100\% ALFALFA ``A-grids" catalog is available at \url{http://egg.astro.cornell.edu/alfalfa/data/.}}, we have access to the {\HI} masses of a large number of galaxies in the direction of the Virgo cluster, and with uniform sensitivity. In order to calculate the {\HI}-to-stellar mass of the reference sample, {\HI} detections from the $\alpha$.100 catalog were cross-matched with the Extended Virgo Cluster Catalog \citep[EVCC;][]{kim14}, within a clustocentric distance of $\sim$3 $\times$ $R_{200}$ \citep{mclaughlin99,ferrarese12}. The lower limit on declination (0$^{\circ}$) comes from the ALFALFA catalog coverage. The upper cut in the radial velocity is 3000~km~s$^{-1}$ \citep[$\sim$43~Mpc assuming $H_{0}$ of 70 km~s$^{-1}$~Mpc$^{-1}$, $\Omega_{\rm M}$=0.3, $\Omega_{\Lambda}$=0.7;][]{wright06}, where a dip in the number distribution of galaxies is present \citep{mei07,kim14}. In total, 580~{\HI}-detected galaxies are cross-matched and available as the reference sample. Stellar masses of the sample were estimated based on the EVCC $(g-i)$ color \citep{zibetti09,kim14}.

By plotting only {\HI}-detected galaxies in phase space, the shape of the very central region of the cluster cannot be seen clearly since most of the virialized galaxies are either early type or have been stripped of their {\HI}, and hence are missed. However, the location of the {\HI} detections in phase space is still intriguing and useful, since important regions for gas stripping and merging structures can be seen.

\section{A phase-space diagram toward the Virgo cluster} \label{sec:phase_gen}

\begin{figure*}
\epsscale{0.8}
\plotone{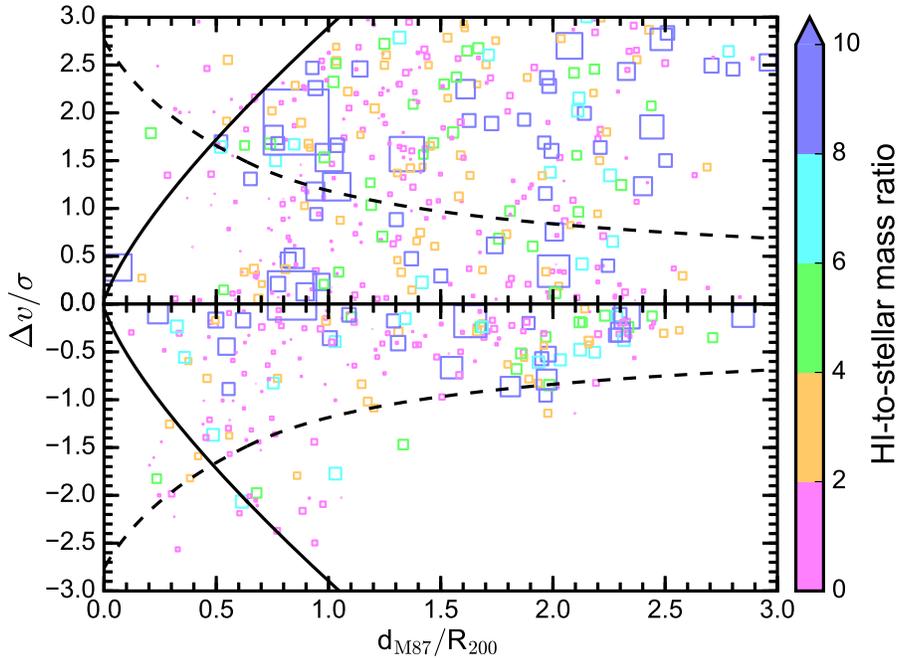}
\caption{ALFALFA reference sample, shown in projected phase space. The color and size of a symbol indicate each galaxy's {\HI}-to-stellar mass ratio. The diagram is unfolded to show the distribution of red-/blueshifted galaxies and substructures, with respect to the cluster mean. The dashed lines represent the escape velocity of the cluster at a particular radius, and the solid lines border the region in phase space where the ram pressure on the {\HI} exceeds the restoring pressure provided by a $\sim$~10$^8$~$M_{\odot}$ galaxy's gravitational potential. Further details on how to calculate the location of these lines can be found in the text. \label{fig:fig3}}
\end{figure*}

A phase-space diagram consists of two observables - the distance from the cluster center and the line-of-sight velocities with respect to the cluster mean. It is called a ``projected" phase space since the clustocentric distance is measured in projection and the velocity along the line of sight. Figure~\ref{fig:fig3} presents the distribution of the reference sample in phase space. The clustocentric distance is normalized by $R_{200}$, and clustocentric velocity has been normalized by the cluster dispersion, $\sigma_{\rm cl}$. We assume that the cluster is centered on the \object{M87} \citep{bohringer94}, and the cluster radial velocity is 1088~km~s$^{-1}$, matching that of the mean radial velocity of the \object{M87} subgroup \citep[Cluster~A;][]{mei07}. Values of $R_{200}=1.55$~Mpc \citep{mclaughlin99,ferrarese12} and $\sigma_{\rm cl}=593$~km~s$^{-1}$ \citep{mei07} are adopted.

Several guide lines are provided in Figure~\ref{fig:fig3} for reference. Dashed lines indicate the escape velocity of Virgo at a given radius, computed assuming that a mass of $4.2 \times 10^{14}$~$M_{\odot}$ is distributed following an NFW potential \citep{nfw96, mclaughlin99}. Solid lines border the region in phase space where the ram pressure exceeds the anchoring pressure, at the center of a galaxy with a mass of $\sim$10$^8$~$M_{\odot}$ (comparable to the median stellar mass of the reference sample). The estimation is based on the following relation, $\rho_{\rm ICM} v_{\rm gal}^2 > 2 \pi G \Sigma_{\rm s} \Sigma_{\rm g}$ \citep[e.g.][]{jaffe15}, where $\rho_{\rm ICM}$ is the ICM density, $v_{\rm gal}$ is the galaxy speed with respect to the ICM, $G$ is the gravitational constant, and $\Sigma_{\rm s}$ and $\Sigma_{\rm g}$ are the stellar and the gas surface densities, respectively.

The efficiency of ram-pressure stripping also depends on the disk inclination relative to the ICM wind direction, in the way that a face-on encounter is more easily stripped \citep[e.g.][]{roediger06, jachym09}, but this effect is not taken into account in our calculation. The central surface gas density was estimated following the prescription of \cite{jaffe15}, adopting a disk scale-length of $\sim$0.5~kpc, the median disk scale-length of the reference sample. The ram pressure was estimated assuming a $\beta$-model density distribution for the ICM of Virgo \citep[$\beta=0.5, r_{\rm c}=13.4$~kpc, $\rho_{0}=4.0 \times 10^{-2}$~cm$^{-3}$;][]{vollmer01}. The complete stripping lines correspond to the projected distance to \object{M87} and the radial galaxy speed component. One should keep in mind that the solid lines are cluster dependent, and also vary from galaxy to galaxy.

\begin{figure*}
\epsscale{0.8}
\plotone{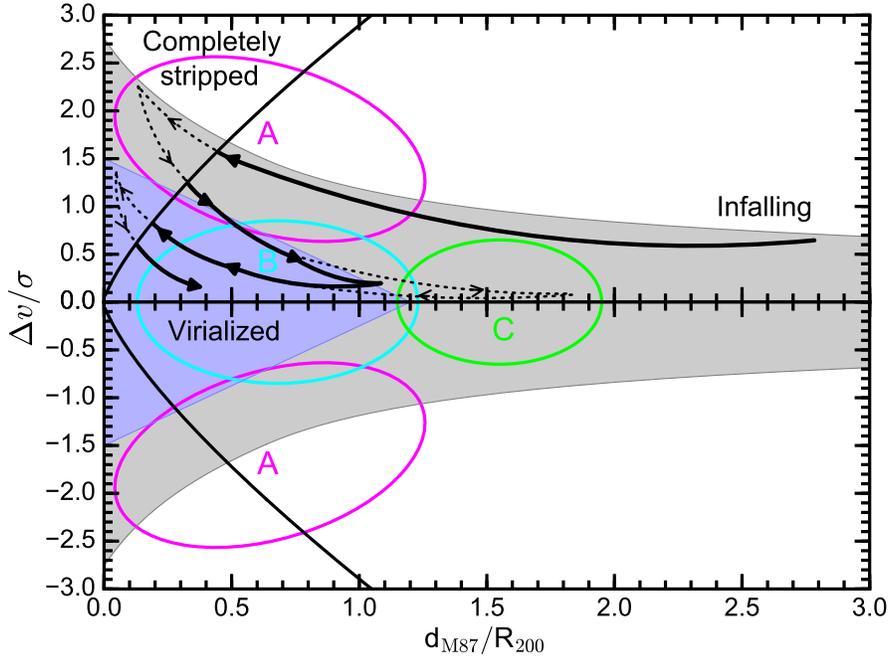}
\caption{Schematic view of the orbit of an infalling galaxy in phase space. The orbital sequence is indicated by lines and arrows (dotted lines for some extreme locations). The area under the influence of the cluster potential is shaded in gray, and the virialized region is shaded in blue. Solid lines have the same meaning as in Figure~\ref{fig:fig3}. For convenience, we roughly divide the cluster into three regions: ({\textsf A}) high velocity at small to intermediate clustocentric distances, ({\textsf B}) low velocity at small to intermediate clustocentric distances, and ({\textsf C}) low velocity at large clustocentric distance. When galaxies fall into the cluster for the first time, they approach from the right side of Region~{\textsf A}. After the first core crossing, galaxies come out from the left side of Region~{\textsf A}, then Region~{\textsf B}, and then backsplash to Region~{\textsf C} (beyond $R_{200}$) in some cases. Then they repeatedly infall and outfall until they settle into the central region of the cluster, while virializing within the potential well. Some galaxies may violently plunge deep inside the cluster, approaching the cluster core quite closely, during the first infall or while virializing.
\label{fig:fig4}}
\end{figure*}

Figure~\ref{fig:fig4} provides a schematic view of the orbit of a galaxy in a cluster showing how it settles or ``virializes" into the potential well as a product of both dynamical friction and violent relaxation. The galaxy's movement is traced by arrows. The gray area represents the region in phase space that is under the escape velocity of the cluster at a particular distance. The virialized region is shaded in blue. The solid lines are the same as in Figure~\ref{fig:fig3}. For convenience, we roughly divide the cluster into three regions: ({\textsf A}) high velocity at small to intermediate clustocentric distances, ({\textsf B}) low velocity at small to intermediate clustocentric distances, and ({\textsf C}) low velocity at large clustocentric distance. 

Once a galaxy is accreted from the cluster outskirts, it radially falls into the cluster with a speed close to the cluster escape velocity or lower, approaching from the right side of {\textsf A} in Figure~\ref{fig:fig4}. As the galaxy approaches to the cluster core, the orbital velocity continues to increase until it reaches its peak value.

In some cases, the galaxy may violently plunge into the cluster, reaching very close to the cluster center, resulting in a very large peak velocity. After the first pericenter, the galaxy moves back toward the cluster outskirts, from the left side of Region~{\textsf A}, passing through the lower part of Region~{\textsf  A} or upper part of Region~{\textsf  B}. Some galaxies may backsplash out to 1.5 to $\sim$2~virial radii, reaching to Region~{\textsf C}. Although it is difficult to assess the orbital/stripping time-scales accurately, a rough estimation of crossing time can still be made. Assuming that a galaxy is moving within a potential of Virgo mass with a mean velocity of 1.2$\sigma$, the time it takes to travel from 2 $\times$ $R_{200}$ to the cluster core ($<$~$\sim$0.5$R_{200}$) is $\sim$3.6~Gyr. To cross the core (the completely stripped zone of $<$~0.5$R_{200}$, or $\sim$0.77~Mpc in Figure~\ref{fig:fig4}), it is estimated to take $\sim$0.6~Gyr assuming a galaxy's speed of 2$\sigma$. This is consistent with the time-scale for galaxies experiencing intensive ram-pressure stripping in a Virgo-like cluster modeled by \citet[][e.g. $200-300$~Myr before and after the peak pressure]{vollmer09}.

As one can expect from Figure~\ref{fig:fig4}, however, the same region may be shared by a number of galaxies at different infalling states. This effect can also be seen in the phase-space probability distribution from the simulations of \cite{oman13}. Therefore, in practice, it is hard to disentangle different populations of observed galaxies in the some critical regions of projected phase space (see Rhee et al. submitted). In this aspect, high-resolution {\HI} data can play a crucial role as it is sensitive to the previous interaction history of galaxies with their surroundings.

Based on this, a few intriguing features can already be found in the phase space of the reference sample shown in Figure~\ref{fig:fig3}. Firstly, the number of {\HI} detections is noticeably small in the stripping zone and the virialized region. Second, quite a number of {\HI}-rich systems are found above the escape line in the top half of the diagram. We confirm that this red-shifted stream of galaxies is consistent with either subgroup M or W \citep{gavazzi99}, which is likely to be either a thin wall behind Virgo, or the \object{NGC 5353}/4 filament, which is a small structure bridging nearby superclusters \citep{kim16}. Considering the velocity range and the distances from \object{M87} of the galaxies, however, the stream in our phase space must be mostly the galaxies in the M$-$W sheet rather than the NGC 5353/4 filament. \cite{kim16} also identified filamentary structures blue-shifted to Virgo, but they are located outside the sky coverage of the EVCC, which explains the absence of galaxies at negative velocities in Figure~\ref{fig:fig3}.

\section{{\HI}-stripped VIVA galaxies in phase space} \label{sec:phase_viva}

The {\HI} morphology, extent, and deficiency of Class~I to $\sim$IV galaxies are highly suggestive of different {\HI}-stripping stages or distinct mechanisms occurring due to the ICM$-$ISM interactions. Therefore, the locations of galaxies in these four classes in phase space can provide insights into the environments where galaxies have lost interstellar gas in the cluster.

In Figure~\ref{fig:fig5a}, \ref{fig:fig5b}, and \ref{fig:fig5c}, the phase-space diagram of the main VIVA targets is shown separated by class and superimposed on top of the ALFALFA reference sample. In order to make the ALFALFA sample more comparable to our VIVA sample, we exclude the galaxies with stellar mass lower than 10$^9$~$M_{\odot}$ from the ALFALFA reference sample that is shown in the background of Figure~\ref{fig:fig5a}, \ref{fig:fig5b}, and \ref{fig:fig5c}. This mass cut also allows us to better separate the cluster and the contaminations.

The influence of Virgo, indicated by the dashed lines, is shown and is identical to that presented in previous figures. The solid lines border zones in which each VIVA galaxy shown in that panel would be completely stripped of {\HI}. These lines indicated the location in phase space where the ram pressure is equal to the anchoring pressure of a galaxy at the galaxy center \citep[$\eta=1$;][]{jaffe15}. The anchoring pressure has been computed for each VIVA galaxy individually, using the observed stellar/gas masses and the effective radius of the sample. It is assumed that the gas mass of a galaxy (before stripping) is 6\% of the stellar mass \citep[for a galaxy with $M_*$ $\sim$10$^{10}$~$M_{\odot}$;][]{gavazzi08}. Once more, we assume a $\beta$-model for the Virgo cluster ICM, with the same parameters as given in Section~\ref{sec:phase_gen}. The ICM density has been taken from the $\beta$-model of Virgo \citep{vollmer01}. Since the galactic dark halo gravitational potential is not included in the calculation of the anchoring force, these lines can be considered as lower limits for complete stripping to occur. We highlight some of the regions where the VIVA galaxies are found using colored ellipses, which match the regions indicated in Figure~\ref{fig:fig4}.

Before discussing the VIVA sample with signs of {\HI} stripping, it is worth inspecting the phase space of Class~0 galaxies presented in Figure~\ref{fig:fig5a}. As expected, galaxies in this class are overall found outside or close to the escape velocity line with higher velocities for a given clustocentric distance compared to the other classes. However, there are several cases sharing the same region of the phase space with {\HI}-stripped galaxies. Those may originate from either the local environmental effects such as tidal interaction and group infall, or the limitations of the projected phase space. Based on the locations of Class~0 galaxies on the phase space, the upper limit on the scatter of $\Delta v/\sigma$ from these two effects (tidal and projection effects) seems to be $\sim$0.5, although it can change depending on the cluster model.

\subsection{Class~I}
During a galaxy's first infall into the cluster, the orbital velocity of the galaxy and the density of the surrounding ICM increase steadily. The galaxy starts to lose gas due to ram-pressure stripping, resulting in an asymmetric {\HI} morphology. But as the ram pressure is still not very strong, there is no truncation of the {\HI} disk yet inside the stellar disk. Thus, the gas/stellar morphologies of the galaxy at the early stage of the first infall resemble that of our Class~I objects.

As shown on the top of Figure~\ref{fig:fig5b}, most galaxies in this class are found in the Region~{\textsf A}. Their {\HI} properties are consistent with a scenario where they are on their way into the cluster center for the first time.

There is one exception, \object{NGC 4698}, which has a very similar velocity to the cluster. However, we note that projection effects could cause the observed velocity to be lower than the true orbital velocity. Similarly, the true 3D radius could be larger than the observed radius. However, when we find multiple objects in the same region of the phase space (e.g. Region~{\textsf A}), the probability that they have similar orbits is significantly larger \citep[e.g.][]{oman13}.

\begin{subfigures}
\begin{figure*}
\epsscale{1.2}
\plotone{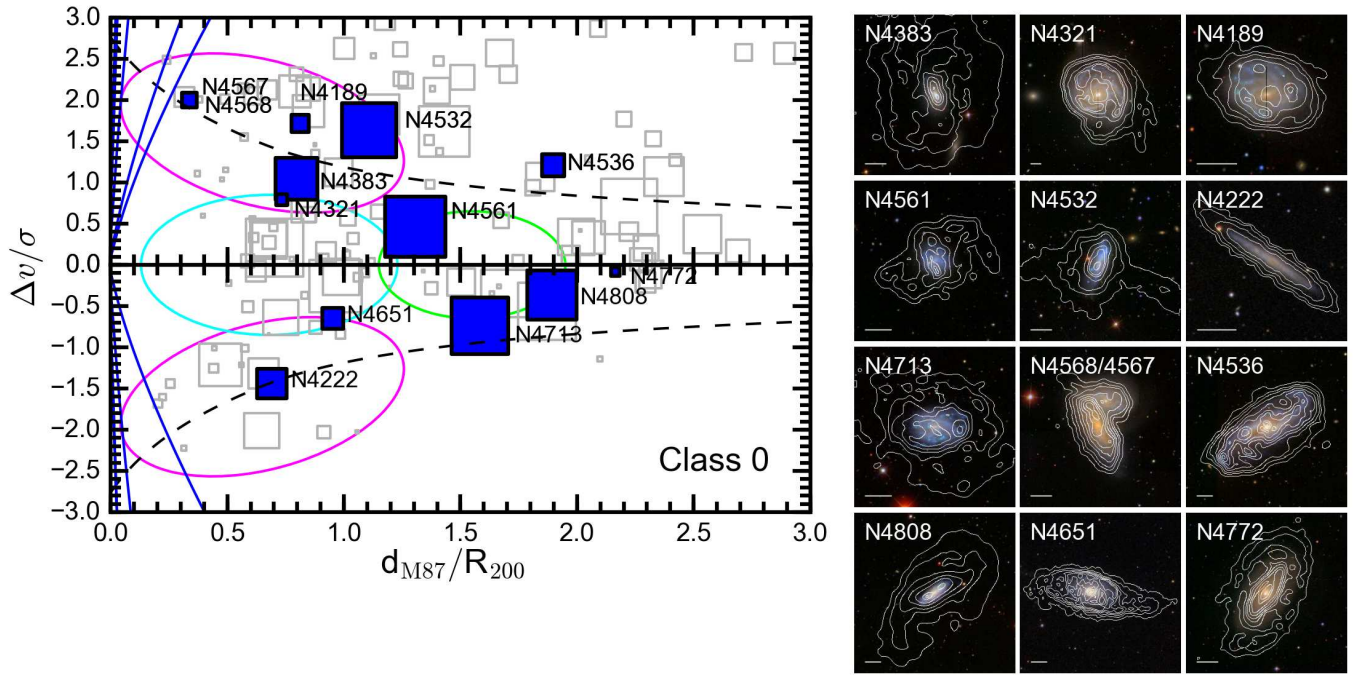}
\caption{Projected phase-space diagram of Class~0 galaxies. The projected clustocentric distance and the line-of-sight velocity with respect to the cluster mean are normalized by $R_{200}$ and the dispersion of Virgo, respectively. The ellipses are the same as in Figure~\ref{fig:fig4}, i.e. notable zones in the orbits of galaxies falling to the cluster, and the dashed line is the same as in Figure~\ref{fig:fig3}, the escape velocity. The solid steep lines indicate where each galaxy shown would be entirely stripped of {\HI}. Galaxies from the ALFALFA reference sample, with $M_{*}$ $> 10^9$~$M_{\odot}$, are shown as open squares in the background, in order to highlight important substructure in phase-space. Galaxy symbols are sized by their {\HI}-to-stellar mass ratios. Contours of {\HI} surface densities \citep{chung09} are overlaid on thumbnail SDSS color images of the galaxies, alongside the phase-space diagrams. The white bar in the bottom-left corner of the thumbnails represents 1~arcmin ($\approx$ 5~kpc) at the Virgo distance. \label{fig:fig5a}}
\end{figure*}

\subsection{Class~II}
As the galaxy approaches to the cluster core, the ram pressure due to the ICM gets stronger, pushing atomic gas away from the galaxy more vigorously. As a result, the galaxy starts to lose the very extended disk gas, and the gas disk is increasingly truncated inside the stellar disk. On the upstream side of this disk, ICM pressure compresses the {\HI}, as seen in Class~II objects.

Galaxies in this class are found in one of two regions in phase space. They are either in Region~{\textsf A} or in Region~{\textsf B}, as shown in the middle panel of Figure~\ref{fig:fig5b}. Region~{\textsf A} is associated with objects that are near first pericenter (either falling in or out of the cluster). Region~{\textsf B} is associated with objects near first apocenter (once more, either falling in or out of the cluster).

Those in the Region~{\textsf A} are likely to be near their first pericenter passage, either right before or immediately after crossing the core for the first time. Among the galaxies found in this zone, two cases are at a large clustocentric distance, \object{NGC 4522} and \object{NGC 4424}. Although \object{NGC 4424} is not located near the core, the ram pressure that can be estimated in a spherical beta model, based on the assumption that the ICM is distributed smoothly, is sufficient to strip the diffuse {\HI} from the outer disk at this location \citep{chung07}. In addition, considering that this galaxy is quite close to \object{M49} in phase space ($d_{\rm M49}$ of 0.45~Mpc and $\Delta v_{\rm M49}=460$~km~s$^{-1}$), it is also possible that this galaxy is one of the galaxies that have recently interacted with the hot gas halo of \object{M49} \citep{biller04}. Meanwhile, \object{NGC 4522} is at the location where the ram pressure is estimated to be too low to explain its {\HI} deficiency and morphology, under the same assumption of smooth and isotropic ICM \citep{kenney04}. However, this galaxy is located near the local X-ray peak, which has been suggested to be responsible for the {\HI} stripping for this case \citep{kenney04}.

Four of the Class~II galaxies are found in the central region of Region~{\textsf B}, which is a transition region. Their locations suggest some galaxies may have long one-sided {\HI} tails or extended features even after passing the cluster core. This could be evidence that the {\HI} streams produced by the earlier cluster passage do not disappear immediately after passing the pericenter. But at the same time, the location of these four cases may suffer from projection effects and/or scatter due to tidal interactions. Intriguingly, all of  these four Class~II galaxies have a close companion (two of which are a pair), which might be responsible for some shift of their locations in phase space as further discussed in Section~\ref{sec:discuss}. Among these four, \object{NGC 4694} is distinct in the sense that its extended tail is actually an {\HI} bridge to its companion. Based on its {\HI} morphology within the disk alone, \object{NGC 4694} is closer to Class~III, which can better explain its location in phase space. It indicates that tidal interactions may not only address a shift of the location in phase space, but also a misclassification in some cases. Among Class~I to IV galaxies, \object{NGC 4694} is the only case connected to a companion through intergalactic {\HI} gas.

\begin{figure*}[h]
\epsscale{1.2}
\plotone{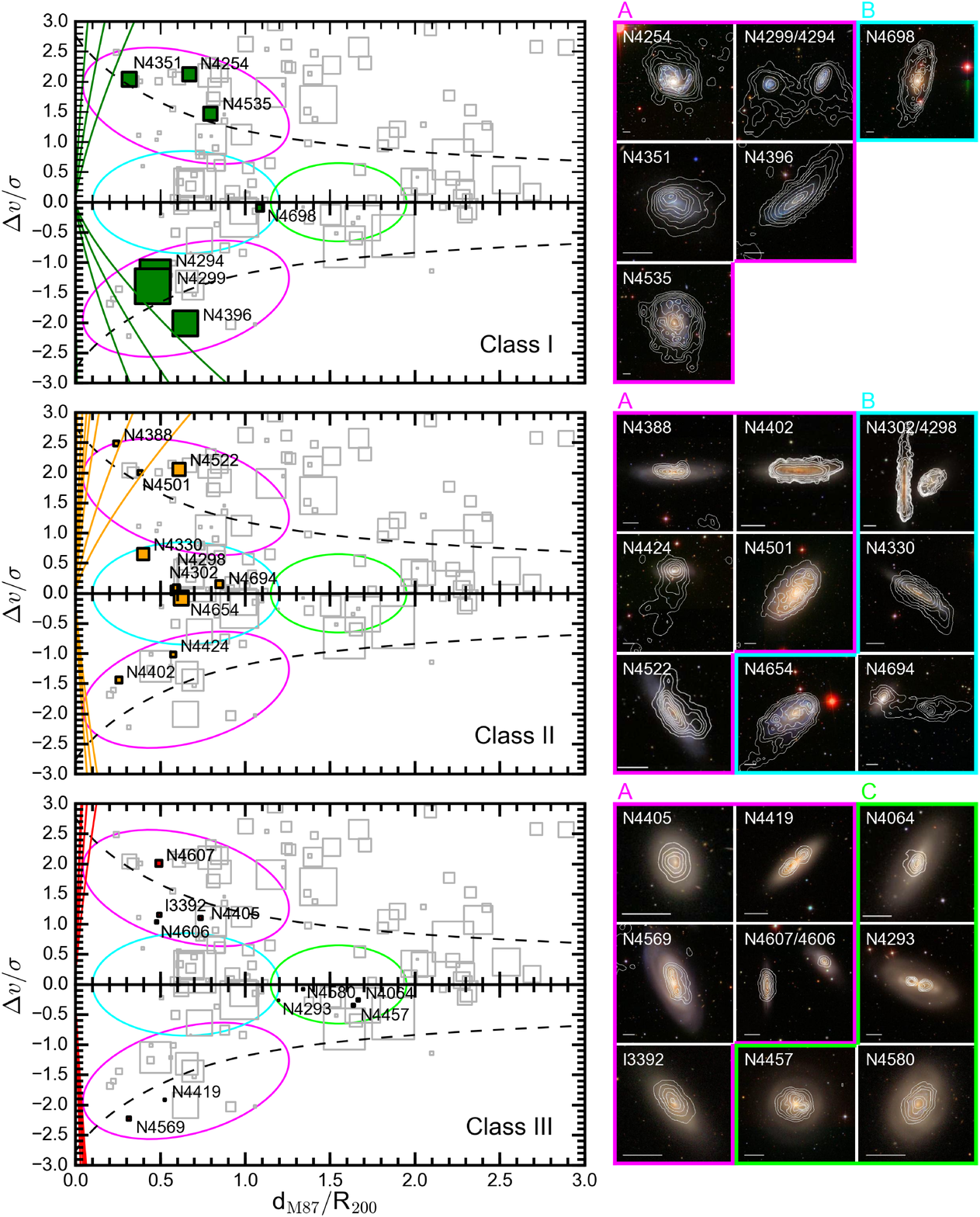}
\caption{Projected phase-space diagrams of Class~I, II, and III galaxies. The thumbnails are also color-highlighted, to match the color of the ellipses shown in the phase-space plot. Details are the same as in Figure~\ref{fig:fig5a}. \label{fig:fig5b}}
\end{figure*}

\begin{figure*}
\epsscale{1.2}
\plotone{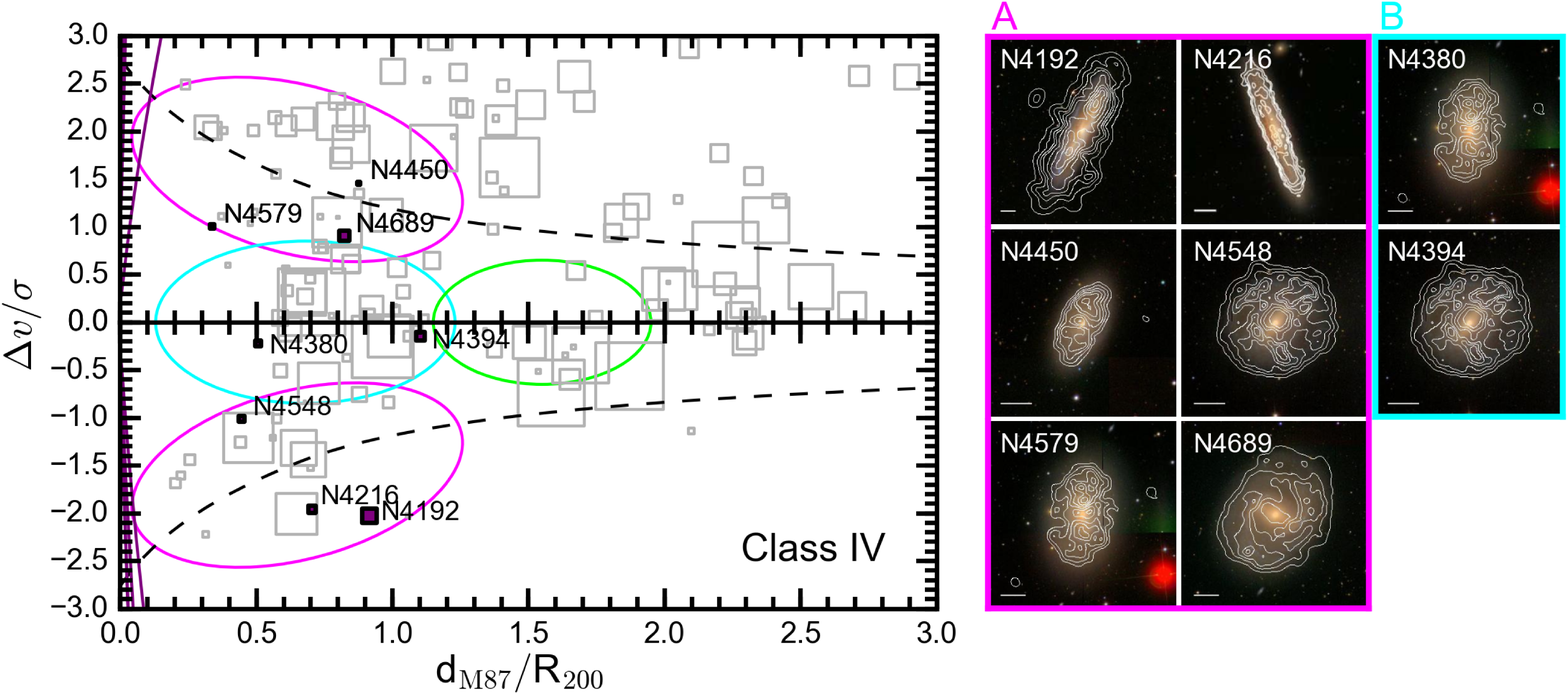}
\caption{A projected phase-space diagram of Class~IV. Details are the same as in Figures~\ref{fig:fig5a} and \ref{fig:fig5b}. \label{fig:fig5c}}
\end{figure*}
\end{subfigures}

\subsection{Class~III}
In some cases, the galaxy may plunge into the cluster more deeply, approaching the cluster center very closely, on their first infall. These galaxies may lose a large fraction of their {\HI} gas due to more radial orbits and hence more extreme clustocentric velocities in the past during the core crossing, resulting in a severely truncated {\HI} disk, as seen in Class~III objects.

Some Class~III objects are found in Region~{\textsf A}, which is associated with being near first pericenter (either falling in or falling out of the cluster), as shown on the bottom of Figure~\ref{fig:fig5b}. However, the heavily truncated {\HI} disks seen in Class~III objects suggest that these objects have already passed pericenter, and therefore falling out is more likely. Furthermore, it is interesting that these objects show no indication of {\HI}-stripped streams, even though they are still close to pericenter, which suggests that the streams were very short-lived. This is in contrast to those Class~II objects that were found in Region~{\textsf B}, with long {\HI} streams. One possibility is that our Class~III objects passed much closer to the cluster center, resulting in heavier disk truncation and reducing the stream lifetime.

This is an interesting example of how phase-space analysis can enrich the information we collect from the {\HI} morphology alone. Based on {\HI} morphology, it might be assumed that Class~III objects always occur after Class~II objects. However, for a small subsample of Class~III objects (those in Region~{\textsf A}), their phase-space location implies that they only just passed pericenter. Meanwhile, for a small subsample of Class~II objects (those found in Region~{\textsf B}), their location in phase space implies they have passed pericenter and are now approaching apocenter. 

After plunging past the cluster core, some galaxies may subsequently reach large radii. Those galaxies, such as those found in Region~{\textsf C}, reach beyond the cluster virial radius. Furthermore, their low velocities with respect to the cluster are exactly as predicted for a backsplashing population of galaxies \citep{gill05}. They can be found out to 2.5 times the cluster virial radius. The galaxies in Region~{\textsf C} may be the clearest examples of backsplashed galaxies found to date. We note that the presence of these galaxies indicates that not all galaxies that have passed pericenter have been converted to passive galaxies, in contradiction to the findings of \cite{oman16}.

\subsection{Class~IV}
Once a galaxy starts to infall deeply into the cluster, it is highly likely to experience ram-pressure stripping. On the other hand, galaxies may lose their gas in other ways, such as starvation due to the stripping of hot halo gas \citep{larson80} or evaporation of gas by thermal conduction \citep{cowie77}. Unlike in ram-pressure stripping, in these mechanisms the gas gets depleted at all radii in the disk and the reduction in {\HI} gas occurs on longer timescales. These more gradual gas removal processes could result in a gas morphology like that seen in Class~IV objects (Figure~\ref{fig:fig5c}).

Thermal evaporation may take place when a galaxy spends sufficient time within the hot cluster medium. Instead of plunging into the cluster along elongated, eccentric orbits, galaxies may follow more circular orbits, which avoid the strongest ram pressure, and so lose their cool interstellar gas by evaporation. However, some studies show that circular orbits are not very probable in the cluster \citep{wetzel11}. Therefore, galaxies showing signs of a slow gas-stripping process may be more likely starved galaxies that have had their halo gas stripped. Stripping of hot halo gas may be caused by various processes \citep{larson80}, and hence those examples can be found at a range of environmental densities. Indeed, Class~IV galaxies are rather spread out in phase space.

\begin{figure*}
\epsscale{0.9}
\plotone{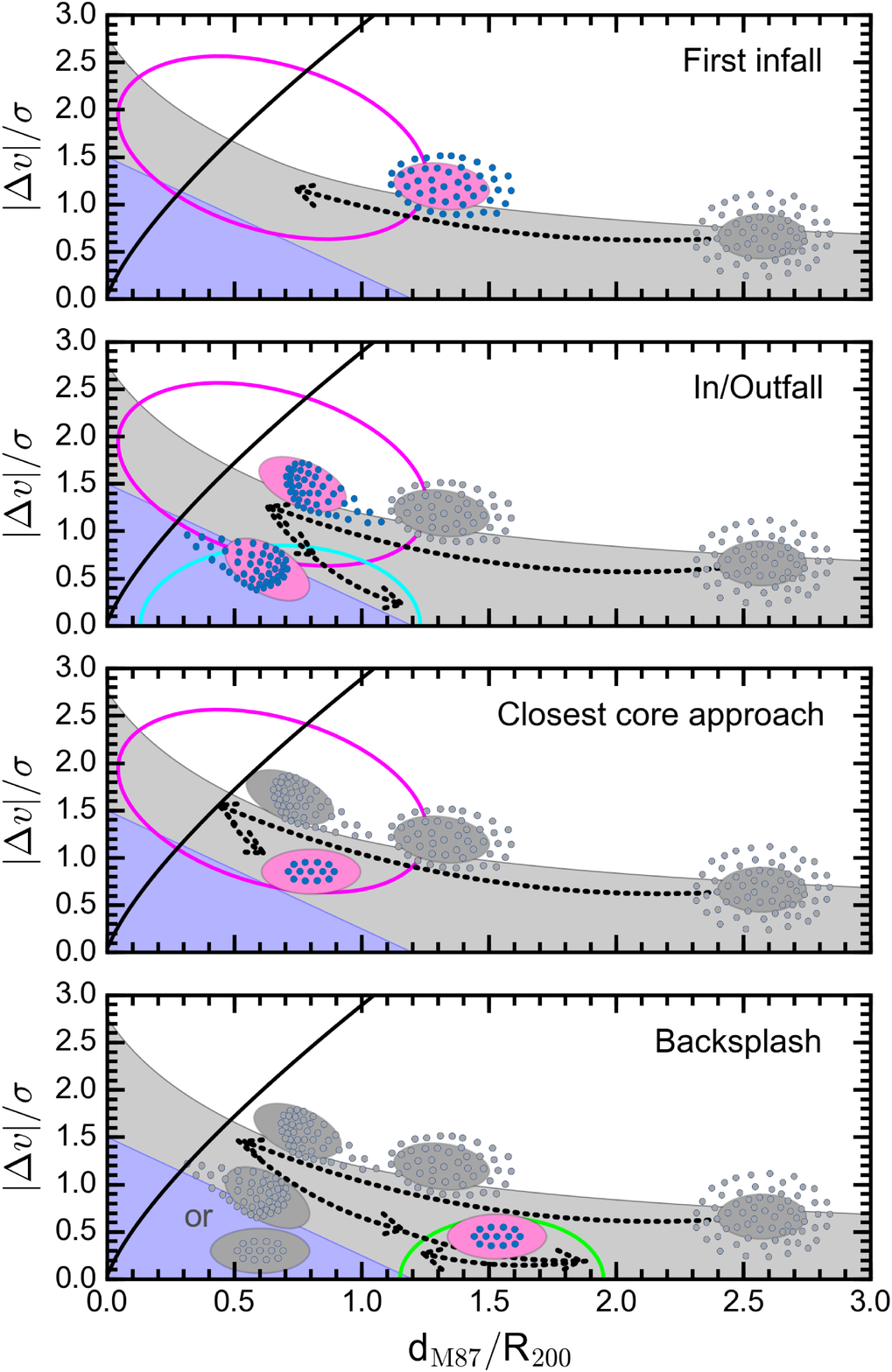}
\caption{Phase-space view of the trajectory of a radially infalling galaxy with illustrations of the {\HI} morphology of galaxies, as part of a time sequence. In each panel, the gray-shaded region represents locations where galaxies are bound to the cluster, and the virialized region is shown in blue. The complete gas-stripping zone for a galaxy with a mass of $\sim$10$^8$~$M_{\odot}$ is shown by a thick black curve, rising from the graph origin. For galaxies, the stellar disk is illustrated with an ellipse, and the {\HI} gas is illustrated by small dots. Galaxies are colored in the positions where the Class~I (top), II (second from the top) and III (third from the top and bottom) galaxies are mostly located.} 
\label{fig:fig6}
\end{figure*}

\begin{figure*}
\epsscale{0.8}
\plotone{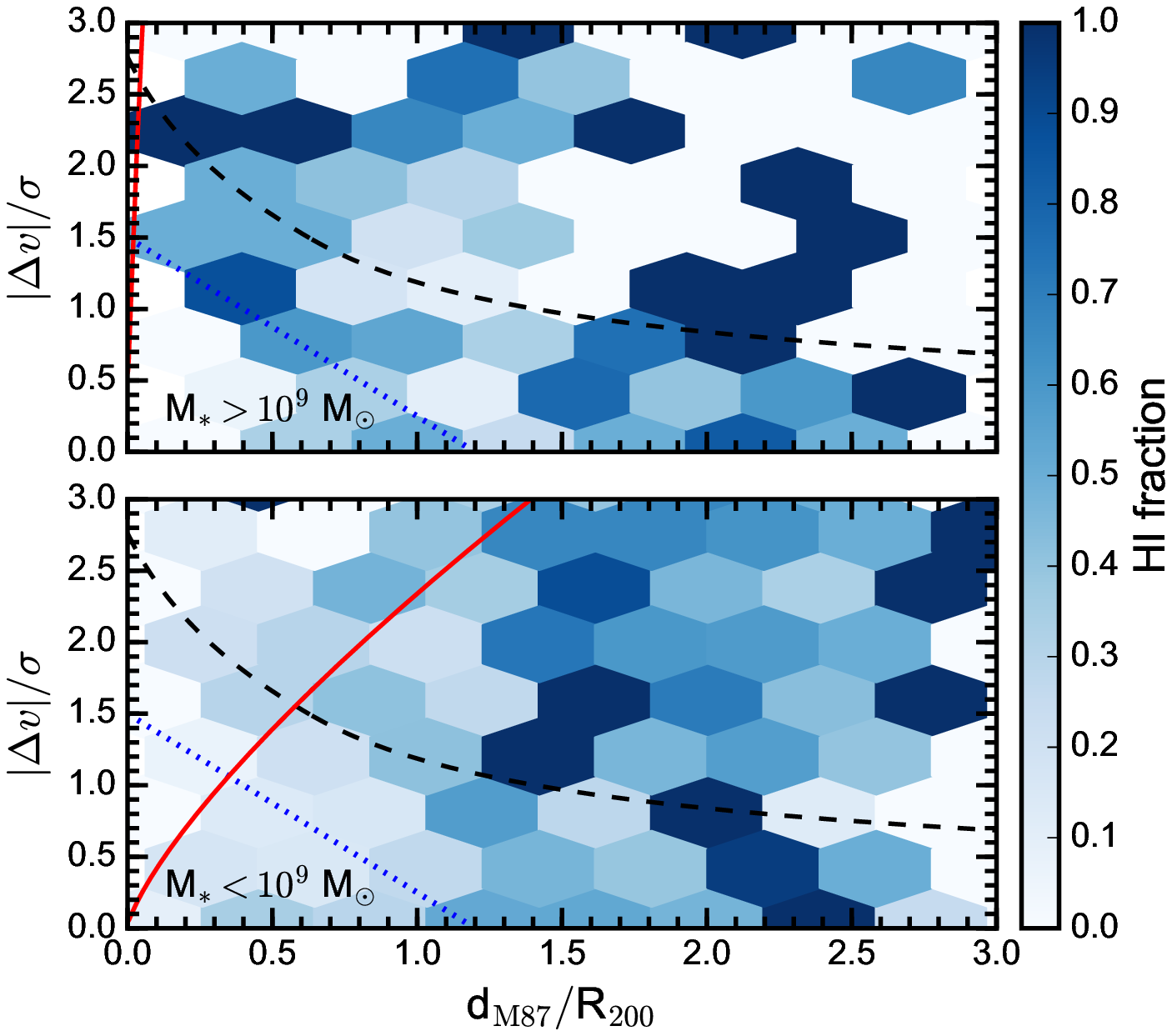}
\caption{Distribution of fractions of {\HI}-detected galaxies in phase space, separated into two subsamples depending on their stellar mass. $M_*$~$>$~10$^9$~$M_{\odot}$ galaxies are in the upper panel, and $M_*$~$<$~10$^9$~$M_{\odot}$ are in the lower panel. The numbers of galaxies plotted are, 327~EVCC galaxies and 131~ALFALFA galaxies for $M_*$~$>$~10$^9$~$M_{\odot}$, and 1259~EVCC galaxies and 449~ALFALFA galaxies for $M_*$~$<$~10$^9$~$M_{\odot}$. Red lines lie on the location in phase space, where galaxies with the median disk scale-length of that mass group would be fully stripped of their {\HI} gas. Blue dotted lines show virialized regions, and the dashed lines indicate the escape velocity of Virgo. \label{fig:fig7}}
\end{figure*}

\section{Discussion} \label{sec:discuss}

\subsection{Trajectories of {\HI}-stripped galaxies in Virgo}
To date, the environment of ram pressure-stripped galaxies has been mostly probed in the perspective of clustocentric distances and/or projected/local X-ray intensities. Meanwhile, tracing the orbital histories of stripped galaxies has largely relied on the use of numerical simulations. Recently, attempts were made to use a phase-space analysis to better understand the origin of the {\HI} deficiencies of galaxies that are embedded in clusters and associated substructures. For example, in the study of a galaxy cluster at $z \sim 0.2$, \object{Abell 963}, \cite{jaffe15} show that the relative {\HI} mass to stellar mass of galaxies can be quite well understood by their positions in phase space under the assumption that ram-pressure stripping is a key mechanism controlling their atomic gas content.

In their study, the phase-space analysis was particularly useful, since the resolution of their {\HI} data is very limited and not sufficient enough to infer the potential trajectory solely from {\HI} morphology. Therefore, the phase space of our sample, with well-resolved {\HI} morphology, can be an important observational confirmation that (1) the positions and the relative {\HI} mass of galaxies are tightly related to their location in their orbit and the status of their gas stripping, as suggested by \cite{jaffe15}, and (2) the classical view of the ram-pressure stripping process works, in practice, in a broad sense, and the role of the ICM for galaxy evolution in the cluster environment is quite significant. 

In Section~\ref{sec:phase_viva}, we have probed the phase-space distribution of galaxies in different phases of {\HI} stripping. In particular, the positions of the Class~I, II, and III galaxies (Figure~\ref{fig:fig5b}), which represent the distinct stages of ram-pressure stripping, and the comparison with Class~0 galaxies (Figure~\ref{fig:fig5a}) can provide the insights into the actual orbits of infalling galaxies.

In Figure~\ref{fig:fig6}, the predicted location of a galaxy that is radially infalling into the cluster while experiencing ram-pressure stripping is shown. While the track in Figure~\ref{fig:fig4} fully follows one galaxy with time, Figure~\ref{fig:fig6} shows ``snap shots," which therefore can be directly compared with our observations, i.e. Figure~\ref{fig:fig5b}. In order to make connections with the infalling history to reach the location where the galaxy is currently found, a potential trajectory is also presented in each panel. From top to bottom, the figure shows the location of a galaxy approaching the pericenter of the first infall (top), in the first in/outfall (second from the top), in the first outfall after a more violent plunge into the cluster core (third from the top), and backsplashed (bottom). 

Figure~\ref{fig:fig6} also illustrates the expected {\HI} morphology of the galaxy at different locations in phase space. The stellar disk is shown by pink ellipses, while the {\HI} gas is plotted with blue dots. The current location at each stripping stage of the galaxy is highlighted in color, while the locations where galaxies used to be in the past are shown in gray. In reality, the gas will be slowed down relative to the ICM after stripping, and therefore one should keep in mind that the direction of the stream does not represent how the stream would lie in phase space, but more how it would appear on the sky.

Generally, there are good agreements between the observations and the predictions, in terms of the positions of the observed galaxies in phase space, and their corresponding {\HI} morphology. First, quite a large fraction of Class~I and II galaxies are found in the ellipses in the top two panels of Figure~\ref{fig:fig6} (85\% and 50\%, respectively). In the case of Class~III, all galaxies are found in the ellipses in the bottom two panels. These results imply that many galaxies experience gas stripping while falling into the cluster along fairly radial orbits, consistent with expectations from cosmological simulations \citep{wetzel10}.

Besides, by combining our sample of VIVA targets with the ALFALFA reference sample, several important features can be seen. One is how {\HI}-detected galaxies differ in location in phase space, as a function of their stellar mass. In Figure~\ref{fig:fig7}, the fraction of {\HI}-detected galaxies, binned into hexagons in phase space, are shown. The upper panel shows the more massive galaxies, with stellar mass $M_* > 10^{9}$~$M_{\odot}$, and the lower panel shows less massive galaxies with $M_{*} < 10^{9}$~$M_{\odot}$. In this region, which is likely under the strong influence of the cluster ($\Delta v/\sigma_{\rm cl} < v_{\rm escape}$ and $d_{\rm M87}/R_{200} < \sim$2), more massive galaxies show a higher probability of being detected in {\HI} than low-mass galaxies. Also, for $M_* > $10$^9$~$M_{\odot}$, many galaxies still contain detectable {\HI}, close to the cluster center. We note that the zone where total removal of the gas is expected (red lines) is much smaller for the higher mass galaxies, as the median galaxy in this sample is more difficult to strip. This confirms the higher anchoring pressure of more massive systems as predicted in Figure~\ref{fig:fig5b} and \ref{fig:fig5c}. 

Also, the fraction of galaxies with {\HI} in the virialized zone near the cluster center is very low for both mass bins, as shown in Figure~\ref{fig:fig7}. This has also been pointed out by \cite{jaffe15}. In their study, which was less sensitive than the ALFALFA survey, a small fraction of the {\HI}-detected galaxies are found in the cluster. In Virgo and its associations, we do find galaxies detected in {\HI} in the virialized zone, perhaps due to a substantially lower {\HI} mass sensitivity limit of ALFLAFA $\alpha$.100  (10$^6$~$M_{\odot}$ at 3$\sigma$), compared to 2 $\times$ 10$^9$~$M_{\odot}$ at 5$\sigma$ in \cite{jaffe15}. Nevertheless, the fraction of galaxies detected becomes much lower in this zone. This indicates that most of the {\HI} is completely stripped from galaxies by the time they are fully virialized within the cluster potential.

Lastly, we notice that some galaxies of the VIVA sample are found outside the escape velocity line, which can originate from several facts. First, this line can vary depending on the details of the model applied. For example, in our model, we assume a spherically symmetric halo. But the halo of real clusters, especially the ones that are in the formation process by accreting substructures like Virgo, is unlikely to be exactly spherically symmetric. Also, we assume that the cluster halo is exactly NFW-like at all radii and is cleanly truncated at the virial radius. In fact, Virgo's outer cluster halo may not be perfectly described by an NFW density distribution and is almost certainly is not cleanly truncated at the virial radius. In fact, observationally, the outer cluster halo is poorly constrained as it is not easily traceable by X-rays, and hence the distribution is highly uncertain. For these reasons, one should take these lines only as an approximate guide.

\begin{figure*}
\epsscale{1.2}
\plotone{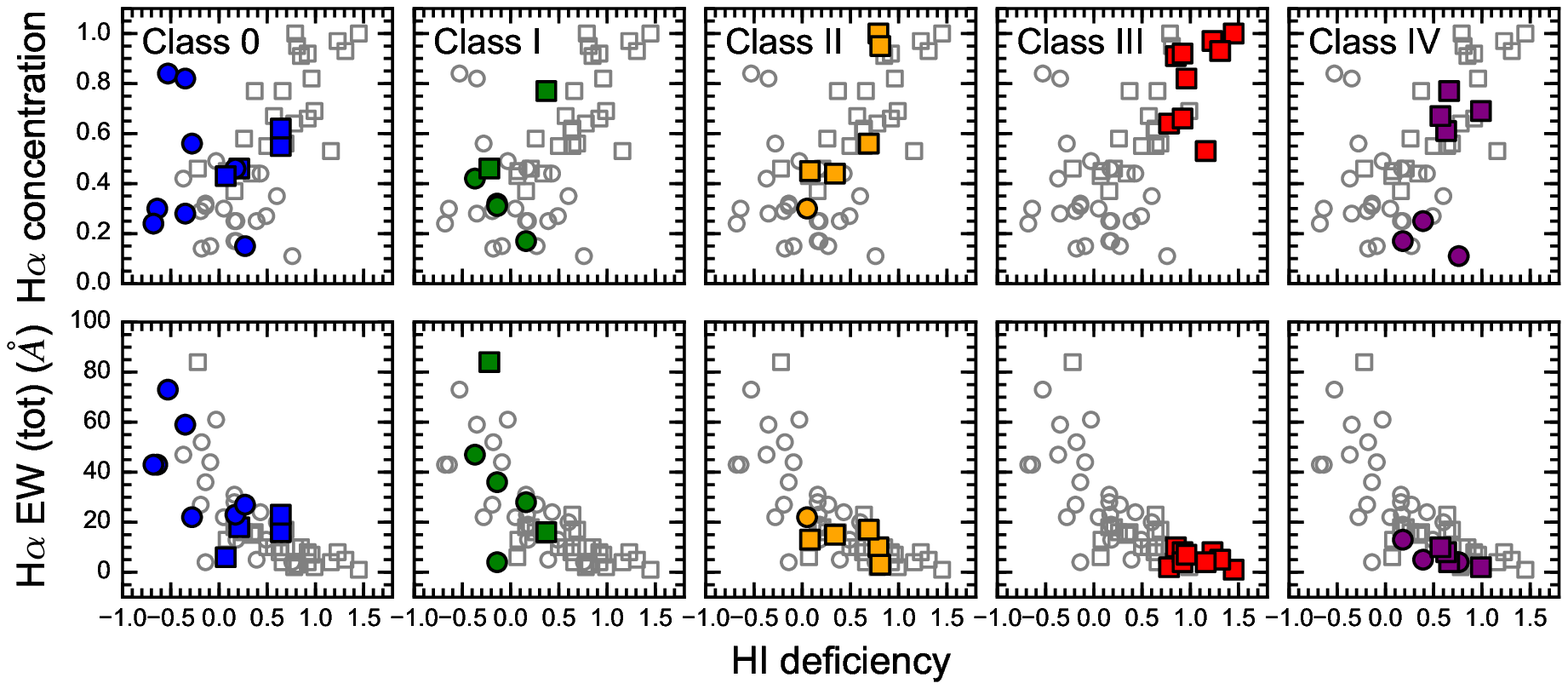}
\caption{{\HI} deficiency vs. H$\alpha$ concentration index ($C$H$\alpha$; top) and total H$\alpha$ equivalent width (bottom) of Virgo spiral galaxies, adopted from \cite{koopmann04b}. In each panel, galaxies of individual subclasses in this study are indicated by a particular color as presented in Figure~\ref{fig:fig1}. Squares are the galaxies with truncated H$\alpha$ disks and circles show no truncations in H$\alpha$.
\label{fig:fig8}}
\end{figure*}

\subsection{Impact of tidal interactions}

In the cluster environment, galaxies experience various tidal processes, such as tidal compression or truncation due to the cluster potential \citep{byrd90,henriksen96}, as well as fast/slow interactions with other galaxies \citep[][p.277]{moore98,mihos04}. In fact, the influence of the cluster potential is difficult to distinguish from the effects of the tidal interactions among galaxies.

However, the gravitational tides generated by the Virgo cluster's mass, estimated based on \citet[][eq. 4-5]{henriksen96}, are found not to exceed the internal gravitational acceleration from \object{NGC 4388} (an MW-sized galaxy), located at a clustocentric distance of 0.37~Mpc, i.e. the closest object in our VIVA subsample to the center of the Virgo cluster. \cite{henriksen96} also reveal that the acceleration due to the cluster potential itself is more important than the drag force within 250~kpc, i.e. at a smaller clustocentric distance compared to the current positions of the VIVA sample for this study. 

However, it is possible that some of our sample have previously gone deeper into the cluster core and been tidally truncated by the cluster potential. If the impact of the cluster potential were noticeably large, backsplashed Class~III galaxies in particular, which are thought to have approached very close to the cluster center, might be expected to be systematically less massive compared to the other classes if they have been tidally stripped of significant stellar mass. But, as seen in Figure~\ref{fig:fig1}, Class~III galaxies are not preferentially smaller in stellar mass. Instead, they are found to have intermediate to high mass among the entire sample, which can be one indication that galaxies are unlikely to have been affected by the cluster potential significantly.

This, however, this does not allow us to completely rule out that those galaxies could have started with a more massive stellar disk. Therefore, we also the inspected morphological properties of the old stellar disk by class based on the $R$-band photometry of Virgo spirals by \cite{koopmann01}. If galaxies had been truncated or compressed, the relative thickness or the compactness might be distinct. However, the relative scale-length to the disk size and the concentration parameter measured in the $R$ band are comparable among all classes. It is quite interesting that Class~III shows a higher fraction of barred galaxies, but it is not significantly high compared to the other classes (80\% for Class~III while 46\%$-$75\% for the other classes), which becomes comparable when only strongly barred cases are counted (40\% for Class~III and 30\%$-$57\% for the other classes). Class~III galaxies also do not appear to be particularly disturbed in morphology compared to the other classes. Therefore, we conclude that the galaxies in our sample have not been significantly affected by the cluster potential, and even if they have been, the impact of cluster potential does not affect the relation between the {\HI} properties and their location in phase space.

Meanwhile, gravitational interactions with other galaxies might lead to a bigger impact on the result. Firstly, regarding galactic morphology, tidal interactions can reshape not only the gas but also the stellar component. But even if the stellar component shows no obvious signs of tidal interactions, the galaxy still could have been tidally affected in several ways. Typically, the {\HI} gas disks extend beyond the stellar disk of a galaxy \citep{warmels88,broeils94}, making the outer disk gas more likely to be tidally stripped than the stars. In addition, the tidal field from neighboring galaxies could soften the potential of a galaxy, making it easier for the gas to be removed \citep[][p.277]{mihos04}. Therefore, both the tidal interactions with companions and the overall tidal fields due to neighbors could make galaxies appear to be experiencing more active gas stripping than might be expected at a given radius, if considering ram-pressure stripping alone. 

Galaxies that show no direct evidence for tidal interactions but are under the influence of surrounding galaxies might result in scatter in associating a galaxy's {\HI} properties with its location in phase space. Among our VIVA sample, \object{NGC 4254}, \object{NGC 4396}, \object{NGC 4294/9} \citep[Class~I;][]{haynes07,chung09}, \object{NGC 4302/4298} \citep[Class~II;][]{chung09}, and potentially \object{NGC 4606/7} (Class~III) could be the examples of the former case. There are several cases that show signs of undergoing minor merging events, \object{NGC 4698} (Class~I), \object{NGC 4424} (Class~II) and \object{NGC 4064} (Class~III), which could potentially have weakened the potential of the host galaxy \citep{cortes06} without destroying its morphology.

Meanwhile, galaxies that are in pairs can move in phase space in both the velocity and radius direction, due to their proper motion. The relative velocity between a pair of massive galaxies with comparable mass can be as large as 500~km~s$^{-1}$ \citep[e.g.][]{casteels13}. Considering the motion of galaxies relative to the center of mass of the pairs, a shift in phase space of up to $\sim\pm$0.4 in $\Delta v$/$\sigma_{\rm cl}$ is possible for Virgo. Candidates for this case among our VIVA sample are \object{NGC 4694}, \object{NGC 4654} \citep[Class~II;][]{vollmer03,duc07}, and \object{NGC 4216} \citep[Class~IV;][]{miskolczi11}. However, the shift in phase space may be considerably lower if the interaction is with lower mass galaxies such as dwarf galaxies, for example, in \object{NGC 4694} and \object{NGC 4654}.

\subsection{Substructures associated with Virgo}

According to the most commonly adopted structure formation theory of the universe \citep[$\Lambda$CDM cosmology;][]{bennett03}, galaxies are accreted into clusters, both individually and also in groups. Also, (sub)clusters of galaxies can merge with one another, forming bigger structures. Therefore, the preprocessing of galaxies, which can affect galactic properties prior to infall \citep{fujita04,mahajan13}, must also be considered in order to understand the location of galaxies in phase space, particularly for a dynamically young cluster like Virgo \citep{gavazzi99,boselli06}.  

Two remarkable subclusters around \object{M49} and \object{M60} \citep[Cluster~B and C;][]{boselli14} are shown in the projected sky of Virgo in X-ray \citep{bohringer94}. According to the averaged radial velocities of these structures \citep[1134~km~s$^{-1}$ for the \object{M49} subgroup and 1073~km~s$^{-1}$ for the \object{M60} subgroup, respectively;][]{boselli14}, they are merging with Virgo from the same plane with respect to the mean of Virgo (1088~km~s$^{-1}$). \object{NGC 4698} in Class~I is located in the \object{M49} subgroup. It suggests that this galaxy might start to lose its gas under the influence of \object{M49} and its substructure. In addition, two cases with outstanding {\HI} tails, \object{NGC 4654} and \object{NGC 4694}, are closer to \object{M60} than to \object{M87}. This suggests that they might be falling into \object{M87}, the main structure of Virgo, as members of a subgroup.

\subsection{Star-formation properties of gas-stripped galaxies}

In this subsection, we discuss how our phase-space analysis can be extended to investigate the evolution of star formation in our galaxies. We find that our {\HI} property-based classes show distinct features in their star-formation. Figure~\ref{fig:fig8} shows good correlations between the {\HI} gas and the star formation properties of our sample. Based on the H$\alpha$ analysis by \cite{koopmann04b}, our sample shows a reduced star formation along with decreasing {\HI} gas content from Classes~I, II, to III. The mean H$\alpha$ equivalent width of Class~I is the highest among the subclasses, but the mean H$\alpha$ concentration, i.e. the ratio of the H$\alpha$ flux within 0.3$r_{24}$ to that within $r_{24}$, is the lowest, suggesting that star formation tends to occur throughout the disk of a galaxy. Unlike Class~I, Class~II shows the increased fraction of galaxies with truncated H$\alpha$ disks. It has a lower H$\alpha$ equivalent width, suggesting a reduced star formation but higher H$\alpha$ concentration compared to Class~I. Most of the Class~III galaxies have truncated H$\alpha$ disks, with high H$\alpha$ concentration and low equivalent width. Class~IV has the highest fraction of anemic galaxies compared to other subclasses. The different H$\alpha$ properties of the subclasses support the fact that our classification based on {\HI} is reliable. Also, it further supports ram-pressure stripping as an important quenching mechanism in the cluster environment. Furthermore, it suggests that our analysis allows us not only to probe the orbital histories of gas-stripped galaxies, but also to probe their star formation while gas is removed. 

\section{Summary and Conclusion} \label{sec:sum}

We have investigated the orbital history of VIVA galaxies using a projected phase-space diagram analysis. Specifically, we consider 35~galaxies that present clear signs of undergoing ram-pressure stripping and are at various stages through the stripping process. The most intriguing results can be summarized as follows.

\begin{itemize}

\item[1.] We first classify galaxies into distinct ram-pressure stripping stages (early, active, and past), based on their {\HI} properties (deficiency, extent, and morphology). We combine this classification with their locations in phase space to better understand the orbits they have undergone to present their current {\HI} properties. We assume that galaxies in Virgo follow a similar distribution of orbits as halos in cosmological simulations of clusters, i.e. in the zone where galaxies near pericenter are found (both before and after passing the cluster core), in the zone where galaxies are found after first pericenter, or in the region where backsplash galaxies are located.

\item[2.] In particular, a sample of galaxies at large clustocentric radii with a velocity similar to the cluster mean are excellent candidates for the ``backsplash" galaxies, which are predicted in cosmological simulations \citep{gill05}. Because they may share the same locations in phase space as infalling galaxies, it is often challenging to distinguish between the two populations in the simulation \citep{oman13}. However, the severely stripped {\HI} disks of our sample at this location clearly indicate that they are not falling into the cluster for the first time, and these may be the clearest examples of backsplashed galaxies found to date.

\item[3.] At small clustocentric radii, combined with low clustocentric velocity, the density of {\HI} detection decreases dramatically, indicating that most galaxies are {\HI} stripped when they settle into the cluster center as virialized. This not only confirms the trend with a clustocentric radius found in a number of nearby clusters \citep{solanes01,boselli06} but also agrees with the findings in a cluster at the $z=0.2$ cluster \citep{jaffe16}. In addition, we confirm that the {\HI} gas in massive galaxies is still detected near the center of the cluster, implying that less massive galaxies get easily stripped in {\HI}.  

\end{itemize}

We note that a projection effect may potentially affect the phase-space analysis. Also, the impact of tidal effects by neighboring galaxies and the subclusters associated with Virgo may change the location of the galaxies in phase space and/or contribute to stripping of {\HI} gas.  However, we confirm that many of our sample, at each stage of gas stripping, are gathered in particular regions in phase space, suggesting that the phase-space diagram is a statistically powerful tool to explain the orbital trajectories of gas-stripped galaxies.

The detailed {\HI} maps of the VIVA survey provide a wealth of information on the environmental effects impacting cluster spirals. We find that by combining them with a phase-space analysis, we can gain additional insight into the orbits they are likely to have followed, the duration for which ram-pressure stripped streams may survive, and we confirm the presence of a clear subsample of backsplashing galaxies. Even at higher redshifts, with poorer spatial resolution of the {\HI}, the relative {\HI} mass (e.g. $M_{\rm\tiny {\HI}}/M_*$) of galaxies is found to be a reasonable first-order indicator of gas-stripping stages. Their locations in phase space can provide a useful way to study not only the accretion history of galaxies around massive structures, but also the evolution of infalling galaxies. 

Lastly, it is also worth mentioning again that the locations of galaxies in phase space, and hence the {\HI} property-based class as we showed, have good correlations with the star formation properties in the way that the fraction of the truncated H$\alpha$ disk dramatically increases from Classes~I, II to III, based on the analysis of \cite{koopmann04b}. In addition, we find the highest fraction of anemic galaxies among Class~IV. That is, the phase-space diagram can be used to probe not only the orbital histories of cluster galaxies that are experiencing gas stripping, but also to probe the impact on their evolution as gas is removed. 

\acknowledgments

We are grateful to the anonymous referee for the constructive comments, which helped us to improve the manuscript tremendously. We thank to the VIVA collaboration for making their fully processed data public. Support for this work was provided by the National Research Foundation of Korea to the Center for Galaxy Evolution Research (No. 2010-0027910) and the Science Fellowship of the POSCO TJ Park Foundation. This work has also been supported by NRF grant No. 2015R1D1A1A01060516 and the Endeavour Research Fellowship of the Australian Government (No. 5030\_2016). This work was co-funded under the Marie Curie Actions of the European Commission (FP7-COFUND). R.S. acknowledges support from the Brain Korea 21 Plus Program (No. 21A20131500002) and the Doyak Grant (No. 2014003730). We also acknowledge to the work of the ALFALFA team and the EVCC team in releasing their useful catalogs. This research has made use of the NASA/IPAC Extragalactic Database (NED; https://ned.ipac.caltech.edu) and the Sloan Digital Sky Survey (SDSS; http://www.sdss.org/).



\end{document}